\documentclass[aps,prx,twocolumn,superscriptaddress]{revtex4-1}
\usepackage{graphicx}
\usepackage{dcolumn}
\usepackage{bm}
\usepackage{color}
\usepackage{ulem}
\usepackage[usenames,dvipsnames,svgnames,table]{xcolor}

\begin{document}




\title{Thermal conductivity of the quantum spin liquid candidate EtMe$_3$Sb[Pd(dmit)$_2$]$_2$: \\
No evidence of mobile gapless excitations}

\author{P.~Bourgeois-Hope}
\affiliation{Institut Quantique, D\'epartement de physique \& RQMP, Universit\'e de Sherbrooke, Sherbrooke, Qu\'ebec, Canada}

\author{F.~Lalibert\'e}
\affiliation{Institut Quantique, D\'epartement de physique \& RQMP, Universit\'e de Sherbrooke, Sherbrooke, Qu\'ebec, Canada}

\author{E.~Lefran\c cois}
\affiliation{Institut Quantique, D\'epartement de physique \& RQMP, Universit\'e de Sherbrooke, Sherbrooke, Qu\'ebec, Canada}

\author{G.~Grissonnanche}
\affiliation{Institut Quantique, D\'epartement de physique \& RQMP, Universit\'e de Sherbrooke, Sherbrooke, Qu\'ebec, Canada}

\author{S.~Ren\'e~de~Cotret}
\affiliation{Institut Quantique, D\'epartement de physique \& RQMP, Universit\'e de Sherbrooke, Sherbrooke, Qu\'ebec, Canada}

\author{R.~Gordon}
\affiliation{Institut Quantique, D\'epartement de physique \& RQMP, Universit\'e de Sherbrooke, Sherbrooke, Qu\'ebec, Canada}

\author{S.~Kitou}
\affiliation{Department of Applied Physics, Nagoya University, Nagoya 464-8603, Japan}

\author{H.~Sawa}
\affiliation{Department of Applied Physics, Nagoya University, Nagoya 464-8603, Japan}

\author{H.~Cui}
\affiliation{RIKEN, Wako-shi, Saitama 351-0198, Japan}

\author{R.~Kato}
\affiliation{RIKEN, Wako-shi, Saitama 351-0198, Japan}

\author{L.~Taillefer}
\email[]{louis.taillefer@usherbrooke.ca}
\affiliation{Institut Quantique, D\'epartement de physique \& RQMP, Universit\'e de Sherbrooke, Sherbrooke, Qu\'ebec, Canada}
\affiliation{Canadian Institute for Advanced Research, Toronto, Ontario, Canada}

\author{N.~Doiron-Leyraud}
\email[]{nicolas.doiron-leyraud@usherbrooke.ca}
\affiliation{Institut Quantique, D\'epartement de physique \& RQMP, Universit\'e de Sherbrooke, Sherbrooke, Qu\'ebec, Canada}

\date{\today}

\begin{abstract}

The thermal conductivity $\kappa$ of the quasi-2D organic spin-liquid candidate EtMe$_3$Sb[Pd(dmit)$_2$]$_2$ (dmit-131) was measured at low temperatures, down to 0.07~K.
We observe a vanishingly small residual linear term $\kappa_0/T$, in $\kappa/T$ vs $T$ as $T \to 0$.
This shows that the low-energy excitations responsible for the sizeable residual linear term $\gamma$ in the specific heat $C$, seen in $C/T$ vs $T$ as $T \to 0$, are localized.
We conclude that there are no mobile gapless excitations in this spin liquid candidate, in contrast with a prior study of dmit-131 that reported a large $\kappa_0/T$ value [Yamashita {\it et al}., Science {\bf 328}, 1246 (2010)].
Our study shows that dmit-131 is in fact similar to $\kappa$-(BEDT-TTF)$_2$Cu$_2$(CN)$_3$, another quasi-2D organic spin-liquid candidate where a vanishingly small $\kappa_0/T$ and a sizeable $\gamma$ are seen.
We attribute heat conduction in these organic insulators without magnetic order to phonons undergoing strong spin-phonon scattering, as observed in several other spin-liquid materials.

\end{abstract}


\maketitle



\section{Introduction}

Low-dimensional materials with triangular or Kagome lattices are fertile playgrounds for the realization of a quantum spin liquid (QSL).
These systems naturally provide a geometrical frustration that suppresses the ordering of antiferromagnetically coupled spins to well below the scale of their coupling $J$, 
leaving the spins in a low-temperature ground state that is highly degenerate and strongly fluctuating, key ingredients of a QSL~\cite{Lee2008,Balents2010}. 
Notable examples of candidate QSL materials are the quasi-2D triangular organic salts $\kappa$-(BEDT-TTF)$_2$Cu$_2$(CN)$_3$ 
(BEDT)~\cite{Shimizu2003} 
and EtMe$_3$Sb[Pd(dmit)$_2$]$_2$ (dmit-131)~\cite{Itou2009}, 
as well as herbertsmithite ZnCu$_3$(OH)$_6$Cl$_2$~\cite{Helton2007,Mendels2007}, with its ideal Kagome lattice. 
In these spin-1/2 systems, early evidence for a QSL ground state came from the absence of magnetic order down to the lowest measured temperature. 
In dmit-131, for instance, an early study reported an exchange interaction $J \simeq 250$~K from susceptibility measurements, 
but no indication of spin ordering or a spin gap from NMR measurements down to 1.4~K~\cite{Itou2009}. 
Instead, below a temperature on the scale of $J$, an increase in antiferromagnetic correlations was seen, a clear effect of the magnetic frustration. 
Similar observations were reported for 
BEDT~\cite{Shimizu2003} and herbertsmithite~\cite{Helton2007,Mendels2007}.

The Holy Grail in the field of quantum spin liquids is to detect mobile gapless excitations, such as `spinons' with fermionic character~\cite{Lee2008,Balents2010}.
The search for thermodynamic and transport signatures of such exotic excitations in a QSL has led to a number of specific heat and thermal conductivity studies. 
The specific heat $C$ of BEDT~\cite{Yamashita2008} and dmit-131~\cite{Yamashita2011} does reveal the presence of low-energy excitations (besides phonons),
down to very low temperature.
When plotted as $C/T$ vs $T$, the data yield a sizeable residual linear term as $T \to 0$ (besides the Schottky term), 
namely $\gamma \simeq 20$~mJ / K$^2$ mol.
Such a residual entropy at the lowest temperatures is a defining property of a spin liquid. 
And indeed, when the same measurement is performed on related organic insulators in which the ground state either has magnetic order, as in $\kappa$-(BEDT-TTF)$_2$Cu[N(CN)$_2]$Cl~\cite{Yamashita2008}, or a nonmagnetic singlet ground state with a large charge gap, as in Et$_2$Me$_2$Sb[Pd(dmit)$_2$]$_2$ (dmit-221)~\cite{Yamashita2011}, one finds $\gamma \simeq 0$.

The thermal conductivity $\kappa$ has an advantage over the specific heat: it is only sensitive to mobile excitations. Localized ones, including the low-temperature nuclear Schottky contribution to specific heat, do not contaminate $\kappa$.
In BEDT, $\kappa(T)$ displays an activated behaviour at low temperature, pointing to gapped mobile excitations~\cite{Yamashita2009}. 
When plotted as $\kappa/T$ vs $T$, the data yield a negligible residual linear term as $T \to 0$, namely $\kappa_0/T  \simeq 0$.
By contrast, a prior study of dmit-131 by the Kyoto group found a large residual linear term $\kappa_0/T  \simeq 2$~mW / K$^2$~cm~\cite{Yamashita2010}.
In a metal, this would correspond to an excellent conductor with a residual resistivity $\rho_0 \simeq12~\mu \Omega$~cm (using the Wiedemann-Franz law, $\kappa_0/T = L_0 / \rho_0$, where $L_0 = 2.44 \times 10^{-8}$~W~$\Omega$/K$^2$).
This is a truly remarkable result. If confirmed, it would be direct evidence of highly mobile gapless (fermionic) excitations in a candidate quantum spin liquid.

In this Article, we report our own thermal conductivity study of dmit-131. 
We observe a vanishingly small residual linear term, such that $\kappa_0/T < 0.02$~mW/K$^2$cm in our 8 samples. 
This is 100 times smaller than the value reported by the Kyoto group~\cite{Yamashita2010}. 
We conclude that the thermal conductivity of dmit-131 does not reproducibly display the signature of mobile gapless excitations.
We attribute heat conduction in dmit-131 entirely to phonons that undergo strong spin-phonon scattering.


\begin{table}[t]
\centering
 \addtolength{\leftskip} {-2cm}
 \addtolength{\rightskip}{-2cm}

\begin{tabular}{|c|c|c|c|c|c|}

\hline
Sample	& Dimensions ($\mu m$)	& Contacts	& a & b & c  \\
		& $l$ $\times$ $w$ $\times$  $t$	&			&    &    & \\
\hline
\hline

 C1  & 400 $\times$ 700  $\times$ 33 & Au pads & -0.016 & 0.42 & 0.75 \\
 C2  & 450 $\times$ 975  $\times$ 41 & Au pads &  -0.020 & 0.33 & 0.58 \\
 C3  & 538 $\times$ 614  $\times$ 35 & Au pads &  -0.025 & 0.23 & 0.61 \\
 D1  & 670 $\times$ 900  $\times$ 60 & GE  &  -0.019 & 0.29 & 0.66 \\
 D2  & 550 $\times$ 1000  $\times$ 60 & Carbon &  -0.019 & 0.19 & 0.69 \\
 E1  & 550 $\times$ 1000  $\times$ 60 & Carbon &  -0.041 & 0.24 & 0.56 \\
 F1  & 500 $\times$ 550  $\times$ 20 & GE  &  -0.012 & 0.30 & 0.74 \\
 G1 & 540 $\times$ 575  $\times$ 15 &  GE & -0.012 & 0.21 & 0.70 \\
 
\hline
\hline
 
A & 1000 $\times$ 1000  $\times$ 50 & Carbon &  1.928 & 57.5 & 2.0 \\
B & 1000 $\times$ 1000  $\times$ 50 & Carbon &  1.095 & 31.5 & 2.0 \\

\hline
\end{tabular}

\caption{List of our 8 measured samples, giving the sample label (where the letter refers to a given crystal growth batch),
the dimensions, the contact method, and the parameters of the power-law fit of the data to $\kappa/T = a + b T^c$ below 0.55~K
(units for $a$ and $b$ are mW/K$^2$cm and mW/K$^{2+c}$cm, respectively).
The combined uncertainty on the sample dimensions gives an overall $\pm$~20\% error on the geometric factor $\alpha = w t / l$.
The last two lines provide the corresponding information for samples A and B of Ref.~\cite{Yamashita2010}, based on their published fits.}
\label{T1}
\end{table}
%
%
%

\section{Methods}

{\it Samples.--}
The crystal structure of dmit-131 EtMe$_3$Sb[Pd(dmit)$_2$]$_2$ can be viewed as a quasi-2D arrangement of alternating Pd(dmit)$_2$ and EtMe$_3$Sb layers. 
EtMe$_3$Sb$^+$ is a non-magnetic closed-shell monovalent cation. The Pd(dmit)$_2$ molecules form dimers each hosting a single spin-1/2 electron, at the vertices of a triangular lattice with slightly anisotropic exchange interactions. 
We measured the in-plane thermal conductivity of 8 single crystals of dmit-131, coming from five different batches (see Table~I). 
These are labelled C1, C2, C3, D1, D2, E1, F1, and G1 where the letters represent a particular growth batch.
All the samples were grown at RIKEN by air oxidation of (EtMe$_3$Sb)$_2$[Pd(dmit)$_2$] (60 mg) in acetone (100 ml) 
containing acetic acid (7-10 ml) at low temperatures in the range of -11 to 5 $^{\circ}$C.
Our sample F1 was prepared in the same conditions, using the same starting material, as the two samples used in the 2010 study by the Kyoto group (labelled A and B)~\cite{Yamashita2010}.
Our sample G1 is coming from the very same growth batch as sample C of Yamashita's recent paper~\cite{Yamashita2019} and as the samples used in specific heat measurements by the Osaka group~\cite{Yamashita2011}.

Scanning electron microscopy measurements on samples C1, D1, D2, and E1, after the thermal conductivity measurements, revealed no signs of micro-cracks down to the sub-$\mu$m level. In addition, single crystal X-ray diffraction (XRD) measurements on C1, D2, F1, and G1 also after our transport measurements, revealed a high crystalline quality as evidenced by a low $R$-factor, typically of the order of 2-4 \%~\cite{Supplementary}, which is comparable to values found in pristine specimens of dmit-131. By all measures, our samples are of comparable structural and crystalline quality as those employed in refs.~\cite{Yamashita2010,Yamashita2019}. Details of our XRD measurements can be found in the Supplementary Material~\cite{Supplementary}.

The heat current was made to flow in the 2D plane of dmit-131.
In samples C1, D2, and G1 the current was along the $a$-axis, and in sample C2 along the $b$-axis. No significant in-plane anisotropy was observed.
The orientation within the plane is not known for our other samples but this is not crucial because of this weak anisotropy (Note that sample F1 is made of two single-crystals with different orientations, see Supplementary Material~\cite{Supplementary}).
The in-plane orientation is not specified in the 2010 study by the Kyoto group~\cite{Yamashita2010}. However, samples C and D of ref.~\cite{Yamashita2019} are $a$-axis samples, as measured by XRD (ref.~\cite{Supplementary}).

To connect wires to the samples, three types of contacts were used:
evaporated gold pads (with silver paint on top) for samples C1, C2, C3;
carbon paste for samples D2, E1;
GE varnish for samples D1, F1, G1.
The leads on all samples were 25~$\mu$m silver wires.
Sample characteristics are given in Table~I.


\begin{figure}[t]
\includegraphics[scale=0.7]{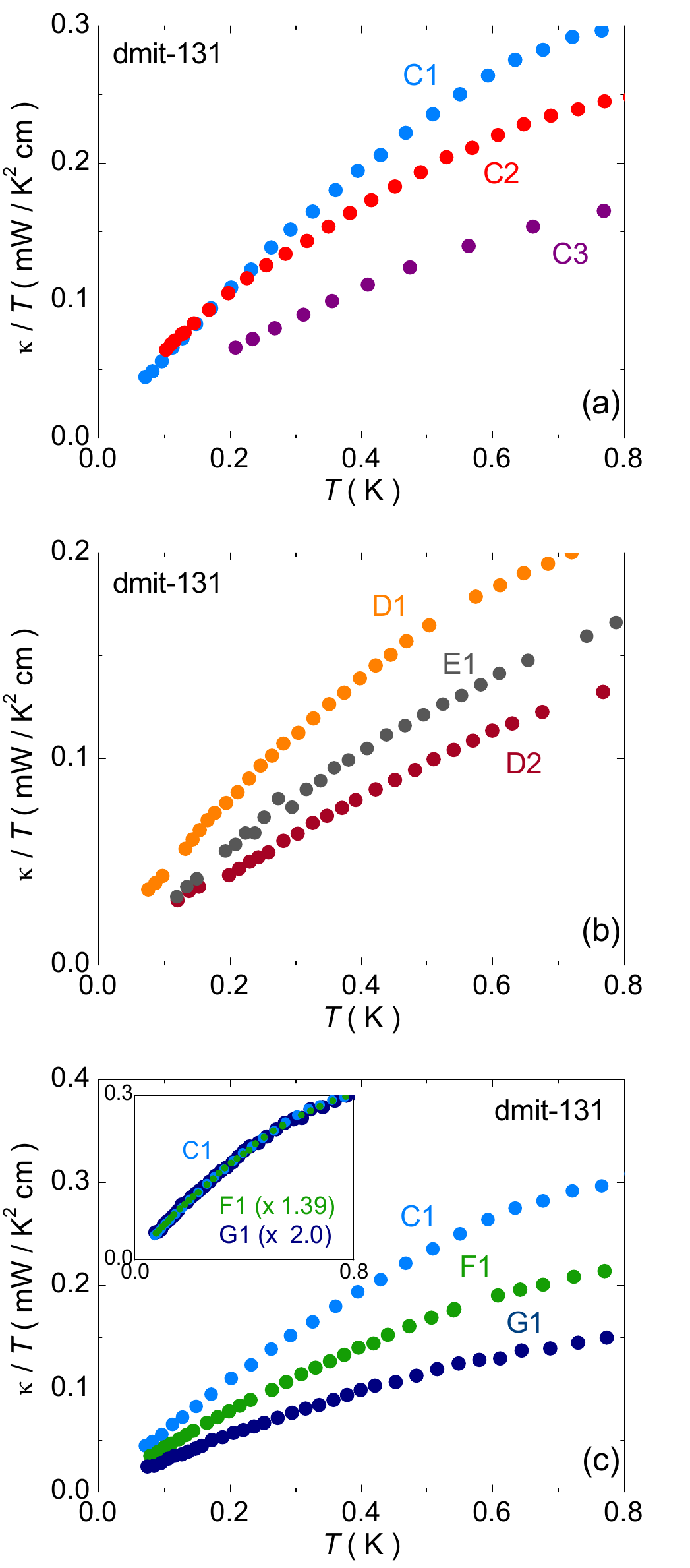}
\caption{
Thermal conductivity $\kappa(T)$ of dmit-131, expressed as $\kappa/T$ vs $T$, 
for samples C1, C2, C3 (a), 
samples D1, D2, E1 (b), and samples C1, F1, G1 (c). 
All data are in zero magnetic field.
In panel (c), the inset compares data for C1 with the data for F1 and G1 multiplied by 1.39 and 2.0, respectively.
}
\label{Fig1}
\end{figure}



\begin{figure}[t]
\includegraphics[scale=0.43]{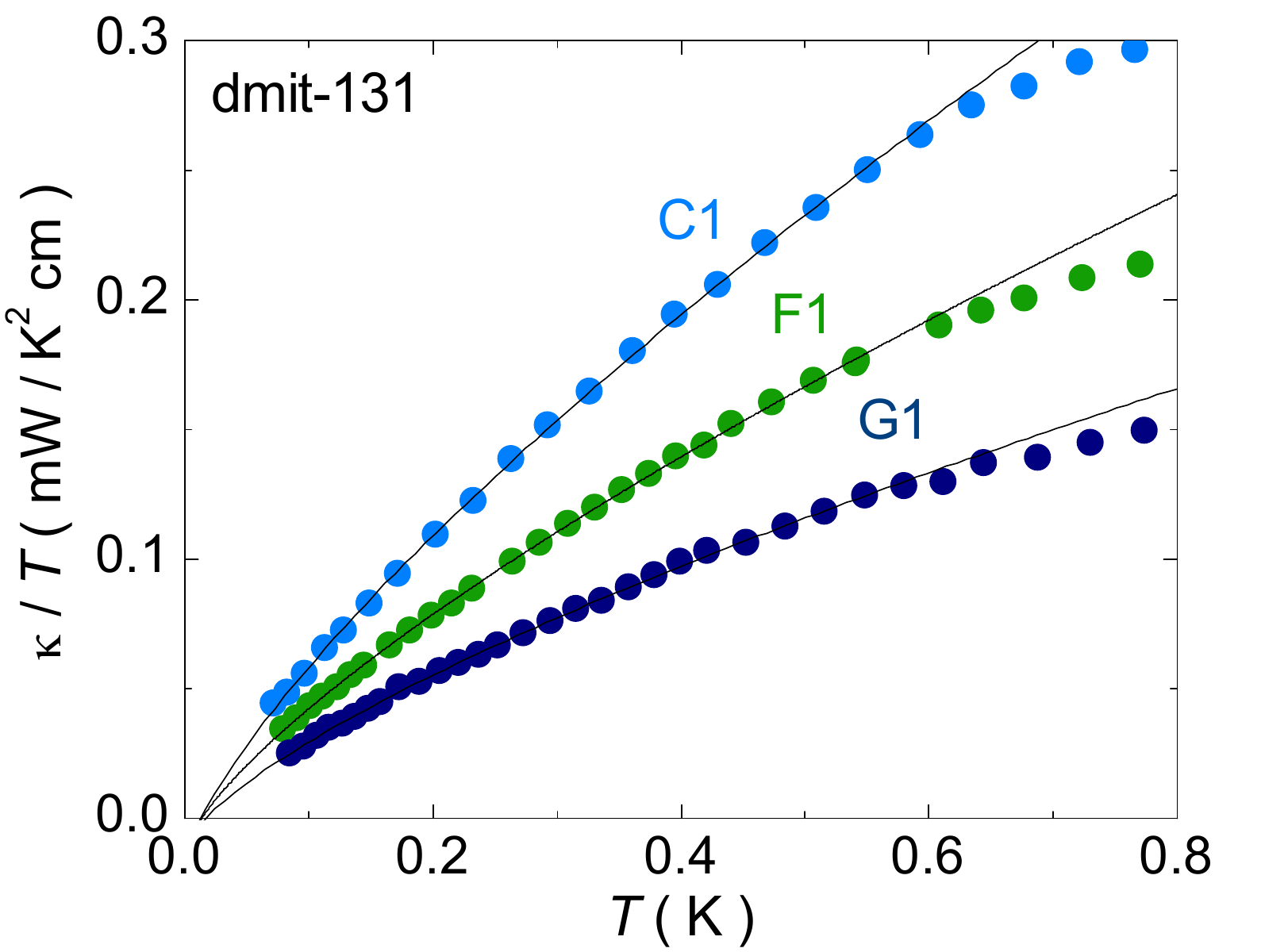}
\caption{
Same data as in Fig.~\ref{Fig1}, for samples C1, F1, and G1. 
The black line is a fit to $\kappa/T = a + b T^c$ below 0.5~K.
The resulting fit parameters $(a, b, c)$ are listed in Table~I for all samples.
}
\label{Fig3}
\end{figure}


{\it Measurement technique.--}
The thermal conductivity was measured using a standard 4-point steady-state method, 
whereby a constant heat current $\dot{Q}$ is injected at one end of a sample whose cold end is connected to a copper block at the reference temperature $T_0$. 
The temperature difference $\Delta T = T^+ - T^-$ is measured at two points along the length of the sample, separated by a distance $L$. 
The thermal conductivity $\kappa$ is given by $\kappa = \dot{Q} / (\Delta T \alpha)$, with the geometric factor $\alpha \equiv A/L$,
where $A$ is the cross-sectional area of the sample (width $w$ $\times$ thickness $t$).
The error bar on the absolute value of $\kappa$ comes mostly from the uncertainty in estimating the geometric factor, approximately $\pm 20$\% for all our samples.

The heat current was generated by sending an electric current through a 10~k$\Omega$ strain gauge whose resistance is independent of temperature and magnetic field. 
The temperatures $T^+$ and $T^-$ were measured using RuOx chips for $T < 4$~K and Cernox sensors for $T > 4$~K, 
calibrated {\it in situ} during the measurement. 
The applied current was chosen such that $\Delta T/T_0 \simeq$~4 - 10\%;
the resulting $\kappa$ was independent of $\Delta T$, indicating that there were no heat losses. 
Measurements below 4~K were performed in a dilution refrigerator, and in a VTI for $T > 4$~K. 
The samples were always cooled slowly from room temperature, at a rate slower than 10~K per hour. 
Tests with a faster cooling rate did not produce a significant difference in the data.

\section{Results}

In Fig.~\ref{Fig1}, we show the thermal conductivity $\kappa$ of all our dmit-131 samples, plotted as $\kappa/T$ vs $T$, in zero field.
We see that all samples display the same qualitative behavior, namely: 
$\kappa/T$ decreases smoothly to zero with decreasing $T$, following a sublinear $T$ dependence.
In Fig.~~\ref{Fig1}(a), we compare samples from the same batch (C), but with different current directions:
along the $a$ axis for C1 and the $b$ axis for C2. 
The shape and magnitude of the $\kappa(T)$ curves are very similar, with no pronounced anisotropy.
Sample C3 from the same batch yields a qualitatively very similar curve, only smaller by a factor 2 or so.
In Fig.~~\ref{Fig1}(b), we compare samples from the same batch (D), but with different types of contact:
GE varnish for D1 and carbon paste for D2. There is no qualitative difference.
In Fig.~~\ref{Fig1}(c), we compare directly samples C1, F1, and G1 which are from different batches, and have different contacts (Au pads for C1, GE varnish for F1 and G1).
In the inset, we see that they have the exact same temperature dependence, {\it modulo} a multiplicative factor of up to 2.0.
As shown in Fig.~\ref{Fig3}, this temperature dependence is well described below 0.5~K by a power-law of the form $\kappa/T = a + b T^c$ where the parameter $a$ is negligibly small, being equal to - 0.016~mW/K$^2$cm for C1, and - 0.012~mW/K$^2$cm for F1 and G1 (values for all our samples are given in Table~I). This shows that $\kappa/T$ extrapolates to a negligible residual linear term in the $T=0$ limit, so that $\kappa_0/T \simeq 0$.
Using instead a linear fit ($c=1.0$) below 0.2~K also yields a negligibly small $a$.
Leaving aside empirical fits, the lowest data point in sample F1, $\kappa/T = 0.03$~mW/K$^2$cm at $T = 75$~mK (Fig.~\ref{Fig3}), imposes an upper bound on the residual linear term of this sample of $\kappa_0/T \simeq0.02$~mW/K$^2$cm.

\begin{figure}[t]
\includegraphics[scale=0.43]{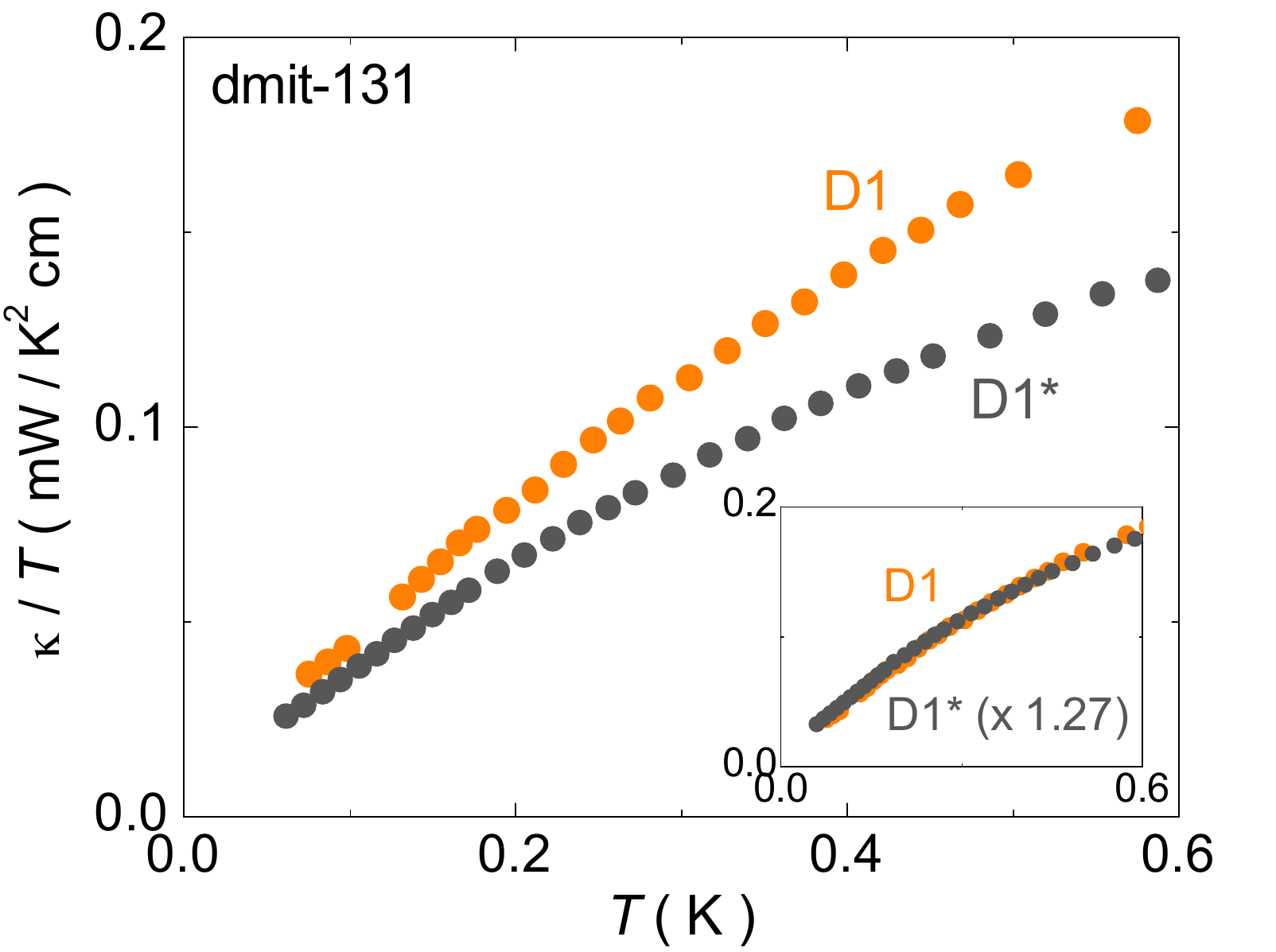}
\caption{
Same data as in Fig.~\ref{Fig1}(b), for sample D1.
D1$^{\star}$ is the same sample, but with contacts rotated in the plane by 90$^{\circ}$ with respect to D1.
Both orientations show the same temperature dependence. 
As seen in the inset, a multiplicative factor of 1.27 collapses the D1$^{\star}$ data onto the D1 data.
This shows that there is no large in-plane anisotropy of $\kappa$ in dmit-131.
}
\label{Fig2}
\end{figure}

Finally, to examine the anisotropy of the in-plane thermal conductivity in a single specimen, D1 was measured with a current along one planar direction, then its contacts were rotated by 90$^{\circ}$ and it was re-measured. As seen in Fig.~\ref{Fig2}, $\kappa$ is very similar for both orientations, {\it modulo} a multiplicative factor of 1.27, which is within the combined uncertainty on the geometric factors.
Considering all the data, we conclude that there is no qualitative difference -- and little quantitative difference -- 
between the five growth batches, the three different types of contacts, and the two current directions, and that $\kappa_0/T \simeq 0$ in dmit-131, with an upper bound of approximately 0.02~mW/K$^2$cm.


\begin{figure}[t]
\includegraphics[scale=0.47]{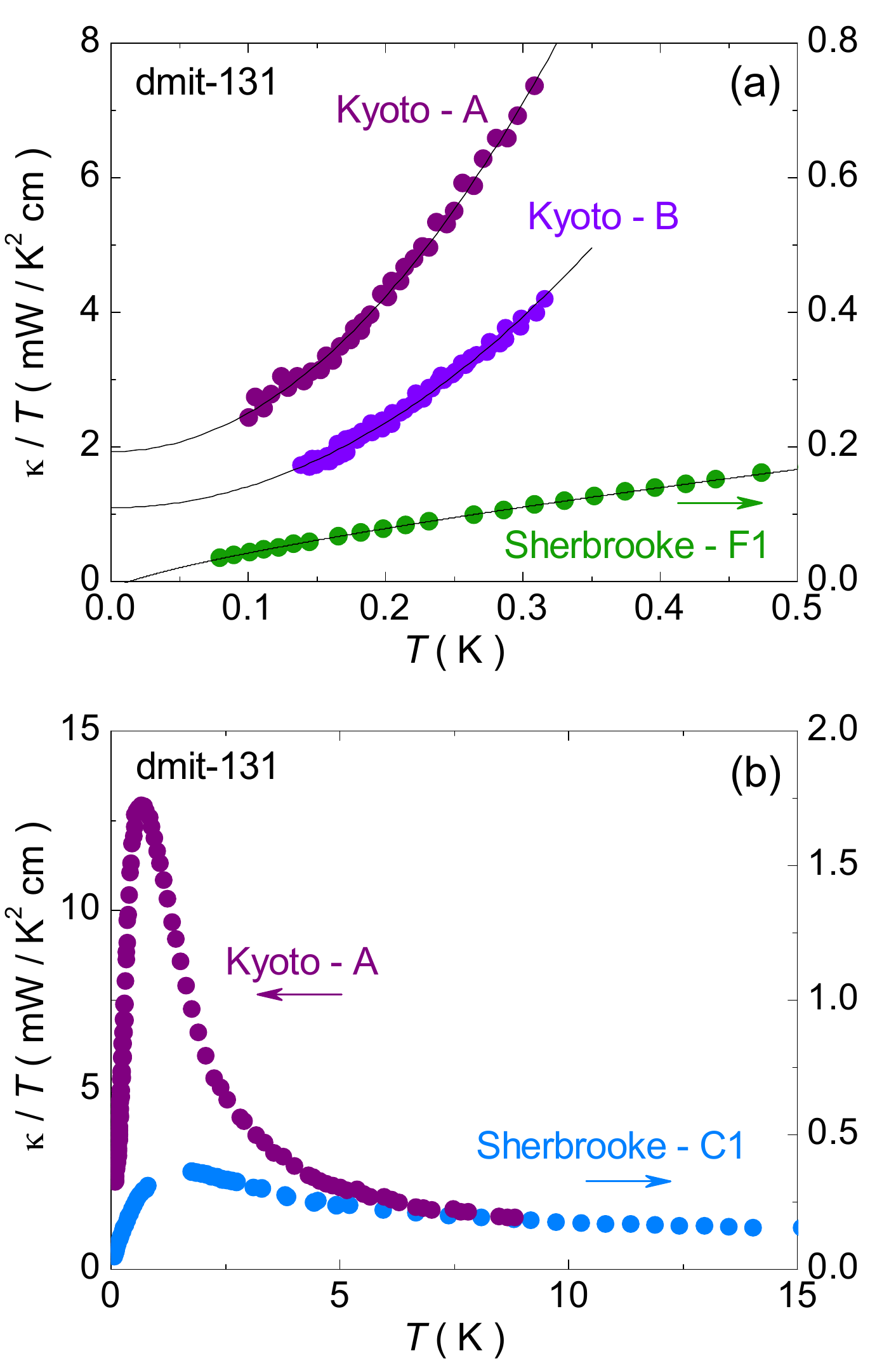}
\caption{
(a)
$\kappa/T$ vs $T$ for our sample F1 (green, right scale)
and for samples A (purple) and B (violet) of ref.~\cite{Yamashita2010} (left scale). 
Note that the left and right vertical scales differ by a factor 10.
The three samples were grown in identical conditions, using the same starting material.
For our data, the black line is the free power law fit below 0.55~K shown in Fig.~\ref{Fig6}(b). 
For the Kyoto data, the lines are quadratic fits reproduced from Ref.~\onlinecite{Yamashita2010}. 
(b) $\kappa/T$ vs $T$ up to 15~K for our sample C1 (blue, right scale) and for sample A of ref.~\cite{Yamashita2010} (purple, left scale).
Note that the left and right vertical scales differ by a factor 7.5.
}
\label{Fig4}
\end{figure}


{\it Comparison to Kyoto data.--}
In Fig.~~\ref{Fig4}, we reproduce the Kyoto data of Yamashita and colleagues~\cite{Yamashita2010} and compare them with ours. 
Note the two distinct $y$-axis scales for their data (left axis) and our data (right axis), which differ by a factor of 
10 and 7.5 in Fig.~\ref{Fig4}(a) and Fig.~\ref{Fig4}(b), respectively.
As readily seen, there is a massive quantitative discrepancy between the
Kyoto data and the Sherbrooke data.
At $T = 0.1$~K, $\kappa/T = 2.5$~mW/K$^2$cm in their sample A (Fig.~~\ref{Fig4}(a)), compared to $\kappa/T = 0.05$~mW/K$^2$cm  in our sample F1 -- a factor 50.
At $T = 0.7$~K, $\kappa/T = 13$~mW/K$^2$cm in their sample A (Fig.~~\ref{Fig4}(b)), compared to $\kappa/T = 0.3$~mW/K$^2$cm  in our sample C1 -- a factor 40.
There is also a qualitative difference between the two data sets.
As seen in Fig.~~\ref{Fig4}(a), the Kyoto data for both of their samples (A and B) show an upward curvature
in $\kappa/T$ vs $T$, whereas all our samples show a downward (sublinear) curvature.
The published power-law fits to their data yield $c \simeq 2 $, whereas we get $c < 1$ (Table~I).
A $T^2$ fit to the Kyoto data yields a huge residual linear term (Fig.~~\ref{Fig4}(a)):
$\kappa_0/T \simeq 1.9$~mW/K$^2$cm in sample A, 
$\kappa_0/T \simeq 1.1$~mW/K$^2$cm in sample B.
This is 50-100 times larger than the upper bound on $\kappa_0/T$ allowed by our data.

\section{Discussion}

Our data on dmit-131 and those of the Kyoto group are in marked contrast. 
We note, however, that their thermal conductivity data on BEDT, another organic quasi-2D insulator with a spin liquid ground state, 
show $\kappa/T$ smoothly going to zero with decreasing $T$~\cite{Yamashita2009}. 
The vanishing of $\kappa/T$ initially follows a sublinear $T$-dependence, as we observe in dmit-131. 
So at very low temperature and zero field, there is no significant difference in the thermal conductivity of the two materials, as indeed there is little difference in their specific heat $C(T)$~\cite{Yamashita2008,Yamashita2011}.

In the absence of a residual linear term $\kappa_0/T$ in the thermal conductivity of dmit-131 and BEDT, 
we conclude that the excitations that give rise to the residual linear term $\gamma$ in the specific heat
of those two materials are not mobile but localized.
In this context, it is difficult to know whether spin excitations make any 
heat-carrying
contribution to $\kappa(T)$.
The conservative view is to assume that all heat conduction is due to phonons,
{\it i.e.} $\kappa = \kappa_{\rm ph}$.
Of course, those localized spin excitations at very low temperature will scatter phonons,
if there is any spin-phonon coupling. 
As it turns out, this scenario is clearly realized in a number of spin-liquid materials.s


\begin{figure}[t!]
\includegraphics[scale=0.42]{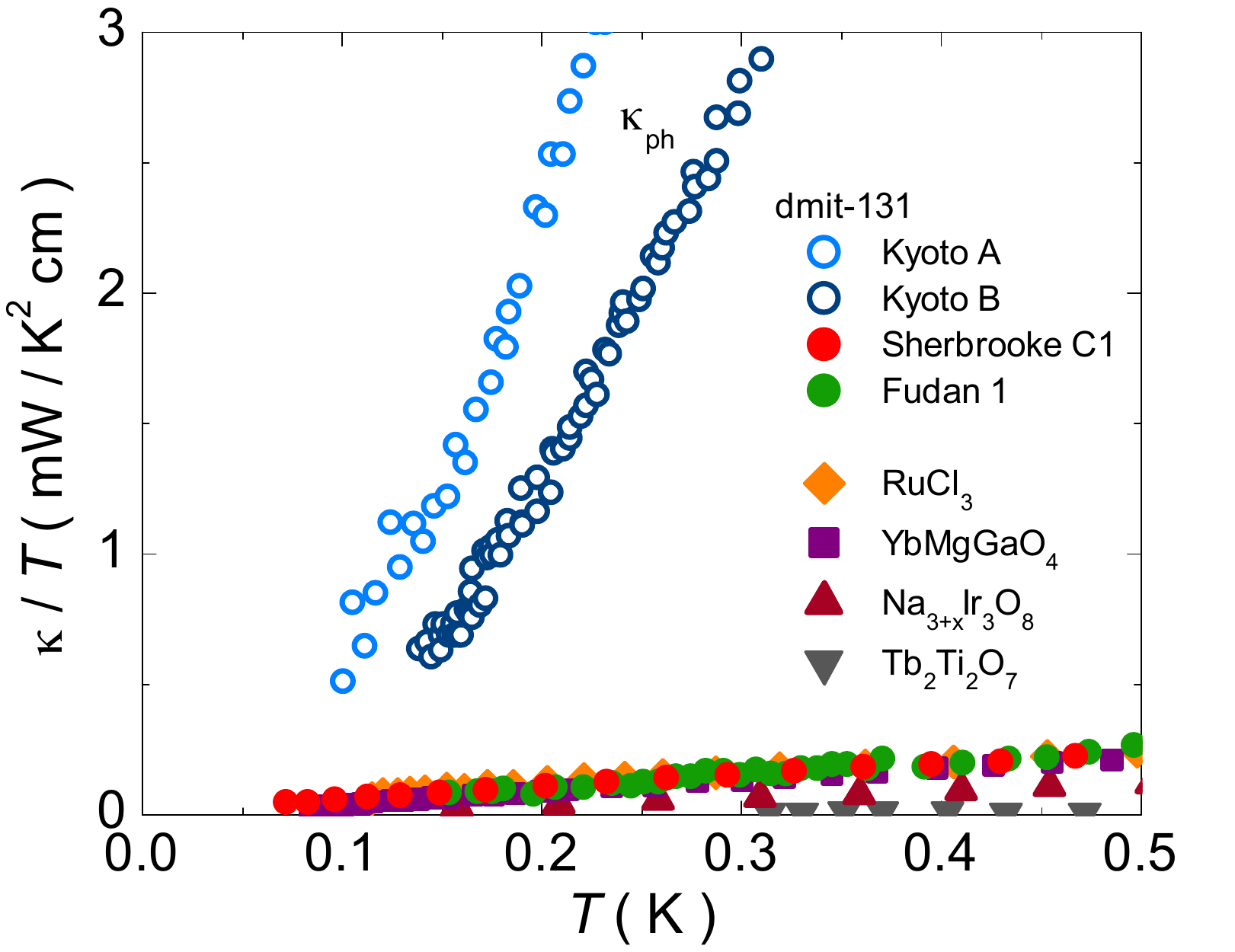}
\caption{
Thermal conductivity $\kappa/T$ vs $T$ for dmit-131, as reported by the Kyoto group~\cite{Yamashita2010}, the Fudan group~\cite{Ni2019}, and in the present work. For the Kyoto data, the residual linear term has been subtracted so that the data reflect only the phonon contribution $\kappa_{\rm ph}$. For comparison, thermal conductivity data for the spin liquid materials RuCl$_3$~\cite{Yu2018}, YbMgGaO$_4$~\cite{Xu2016}, Na$_{3+x}$Ir$_3$O$_8$~\cite{Fauque2015}, and Tb$_2$Ti$_2$O$_7$~\cite{Li2013} are also plotted. These are all insulators so the thermal conductivity is entirely phononic and is at least 20 times smaller than $\kappa_{\rm ph}$ in the Kyoto data.
}
\label{FigA}
\end{figure}


The pyrochlore material Tb$_2$Ti$_2$O$_7$ does not show any magnetic order down to the lowest temperature~\cite{Gardner1999}, as a result of geometrical frustration, reaching instead
a highly degenerate spin-ice state in the ground state.
Owing to spin-phonon scattering, the thermal conductivity of Tb$_2$Ti$_2$O$_7$ is massively suppressed relative to its isostructural
non-magnetic analog Y$_2$Ti$_2$O$_7$, a simple insulator without magnetic moments.
At $T = 0.3$~K, the phonon mean free path in Y$_2$Ti$_2$O$_7$ is set by the sample dimensions, when phonon motion becomes limited by boundary scattering.
By contrast, in Tb$_2$Ti$_2$O$_7$, the phonon mean free path at $T = 0.3$~K is 200 times smaller~\cite{Li2013},
because phonons are strongly scattered by spins in the spin-ice state.

Another example of spin-phonon scattering is provided by the hyperkagome antiferromagnet Na$_{3+x}$Ir$_3$O$_8$,
which can be tuned from a spin-liquid-like insulator to a metal by reducing the Na content~\cite{Fauque2015}.
In the insulator, the scattering of phonons off the gapless magnetic excitations (detected in the low-temperature specific heat)
causes a reduction in $\kappa_{\rm ph}$ by a factor 25 at $T = 1$~K relative to the metal~\cite{Fauque2015}.

Spin-phonon scattering is also clearly observed in YbMgGaO$_4$~\cite{Xu2016} and RuCl$_3$~\cite{Yu2018}.
The spin liquid candidate YbMgGaO$_4$ displays strong magnetic excitations visible as a power-law dependence of the specific heat, but its thermal conductivity is dominated by phonons that are scattered by these excitations~\cite{Xu2016}.
RuCl$_3$, on the other hand, has recently attracted considerable attention as a Kitaev honeycomb magnet. When a field fully suppresses the zigzag magnetic order at about 7.5~T, the phonon-dominated thermal conductivity of RuCl$_3$ reaches a minimum because of scattering by spin excitations~\cite{Yu2018}.

In Fig.~\ref{FigA} we compare the thermal conductivity of these four materials with that of dmit-131 as reported by the Kyoto group~\cite{Yamashita2010}, the Fudan group~\cite{Ni2019}, and ourselves.
We see that our data and those of the Fudan group agree well with those typical of spin-liquid materials, but that the Kyoto data are anomalously large, by a factor of 20 or more.
This reveals a fundamental problem with the Kyoto data: their phonon conductivity $\kappa_{\rm ph}$ is enormous (see samples A, B, C and D of ref.~\cite{Yamashita2019}).
It corresponds roughly to the value expected if the phonon mean free path is the size of the sample (Fig. 2 in ref~\cite{Yamashita2019}).
This would mean that nothing scatters phonons at low temperature, which is physically unrealistic.
The fundamental property of a gapless spin liquid is the presence of zero-energy spin excitations, that show up in a finite residual term $\gamma$ in the specific heat. Those spin excitations will scatter phonons down to very low $T$ – much like electrons in a metal scatter phonons down to very low $T$. We know of no gapless spin liquid candidate whose $\kappa_{\rm ph}$ is not strongly reduced from its maximal boundary-limited value.

\begin{figure}[t!]
\includegraphics[scale=0.43]{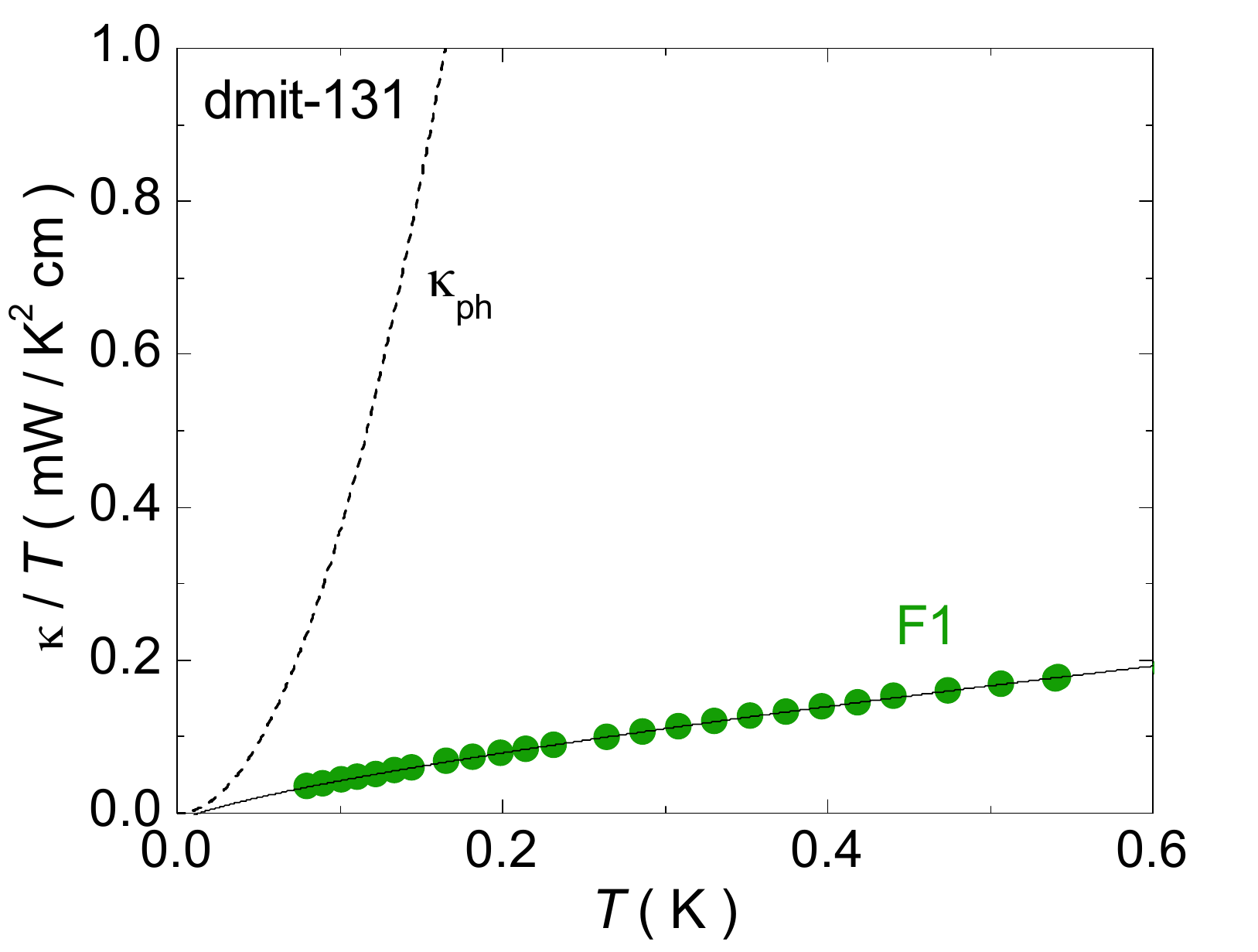}
\caption{
Same data (green dots) and fit (full line) as in Fig.~\ref{Fig3}, for sample F1. For comparison, the dashed line is the phonon thermal conductivity expected for that sample if phonons were only scattered by the sample boundaries
(see text).
}
\label{Fig5}
\end{figure}

We therefore propose that phonons in dmit-131 (and BEDT) are also scattered by the low-energy spin excitations
of their spin liquid state at low temperature.
In Fig.~\ref{Fig5}, we compare our data on sample F1 to the calculated value of $\kappa_{\rm ph}$ assuming the phonon mean free path $\ell_{\rm ph}$ is limited only by the sample boundaries,
so that $\ell_{\rm ph} = 2\sqrt{wt/\pi} = 120~\mu$m (Table~I),
using the standard formula, $\kappa_{\rm ph} = C_{\rm ph} v_{\rm ph} \ell_{\rm ph}/ 3$~\cite{Li2005}, where $C_{\rm ph} = \beta T^3$ is the phonon specific heat
of dmit-131, with $\beta = 24$~mJ/K$^4$mol (47~J/K$^4$m$^3$ given 4 units of chemical formula per unit cell)~\cite{Yamashita2011}, and $v_{\rm ph}$ is the phonon velocity, taken to be 2000~m/s.
We find $\kappa_{\rm ph}$~=~37~$T^3$~mW/Kcm in the boundary scattering limit which, at $T = 0.3$~K, is roughly 30 times larger than the measured value.


\begin{figure}[t!]
\includegraphics[scale=0.43]{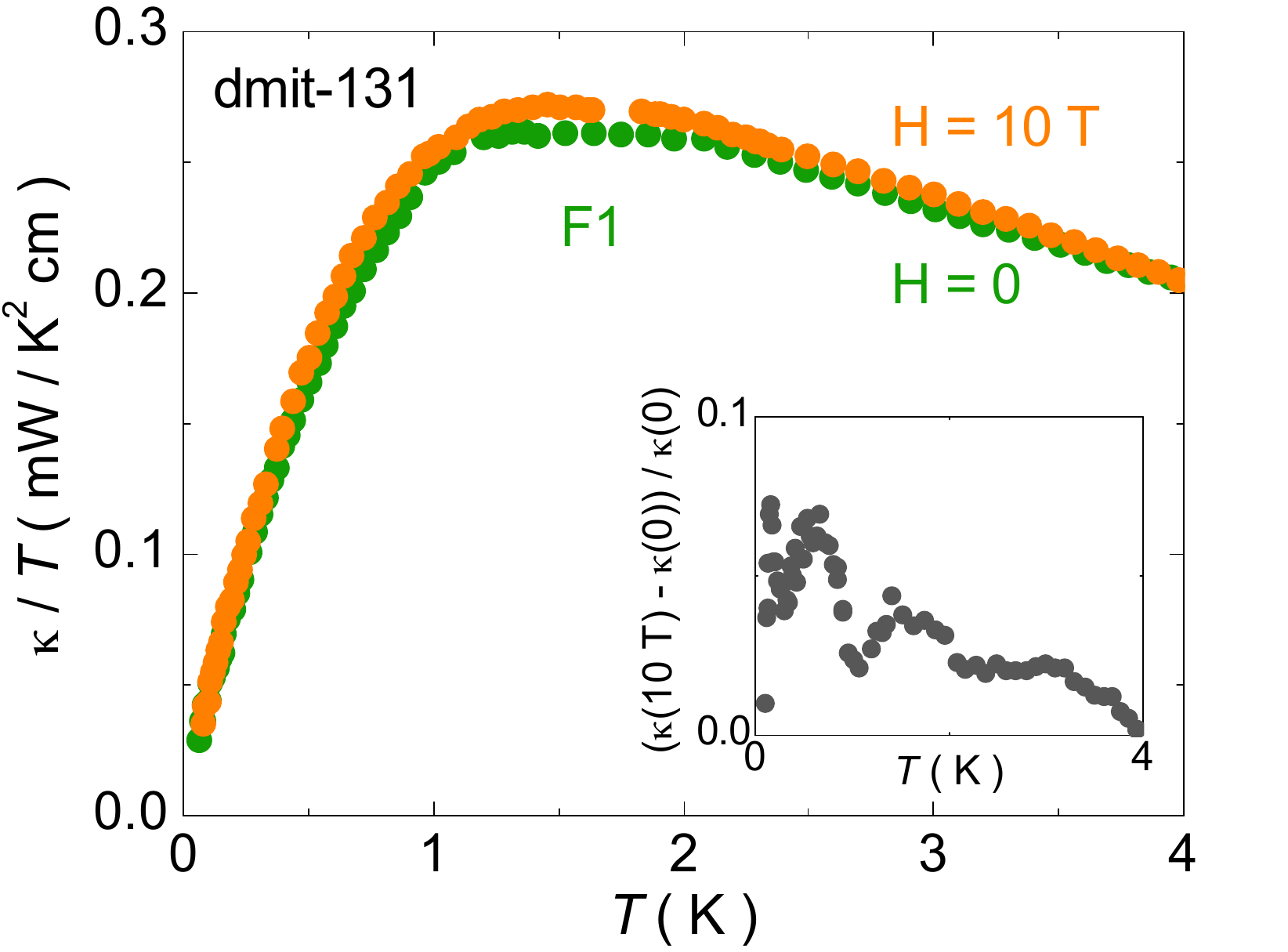}
\caption{
Thermal conductivity of our sample F1, plotted as $\kappa/T$  vs $T$, in both zero field (green) and at $H = 10$~T (orange), revealing a very weak field dependence up to 4~K.
Inset: relative change in $\kappa(T)$ due to a magnetic field $H = 10$~T, plotted as $(\kappa(10~{\rm T}) - \kappa(0)) / \kappa(0)$ vs $T$.
}
\label{Fig6}
\end{figure}

In a scenario where phonons are scattered by spin excitations, one might expect the spin spectrum to be dependent on magnetic field,
and thus $\kappa_{\rm ph}$ to be field dependent.
In Fig.~\ref{Fig6}, 
we see that applying 10~T normal to the 2D planes does increase $\kappa$ in dmit-131, but only very slightly, by no more than 5\% below 4~K.
The weakness of the effect is consistent with the effect of field on the specific heat of dmit-131, which is no more  than 5\% below 4~K~\cite{Yamashita2011}.
Not shown here, our data for a field of 10~T applied in the plane are essentially identical to those shown in Fig.~\ref{Fig6}.


A possible explanation for the unusual low-temperature behavior of both $C(T)$ and $\kappa(T)$ in dmit-131 (and BEDT) is provided by a recent theory of
local singlets in a quantum paramagnet with frustration, that appear as defects in a triangular lattice of spin-1/2 moments~\cite{Kimchi2018}.
These local singlets give rise to a specific heat that grows as $C \propto T^{0.7}$ from $T=0$ and 
they scatter phonons in such a way that $\kappa_{\rm ph} \propto T^2$.
Another scenario is based on a $S$ = 1/2 Heisenberg antiferromagnet on a triangular lattice with quenched randomness in the exchange interaction~\cite{Watanabe2014}, whose quantum spin-liquid ground state has gapless excitations that give $C \propto T$.

Finally, we note that nuclear magnetic resonance measurements on dmit-131~\cite{Itou2009,Itou2011} reveal a vanishing dynamical susceptibility at low temperatures, which excludes fully gapless fermionic magnetic excitations in the ground state of dmit-131. This is entirely consistent with our finding that no gapless fermionic excitations contribute to the heat transport, but hard to reconcile with the conclusion of Yamashita and colleagues~\cite{Yamashita2010}. This shows that the physics of dmit-131 must be seriously re-examined.

\section{Summary}

We measured the thermal conductivity of 8 single crystals of dmit-131 at temperatures down to 70~mK,
using three different types of contact, and find that all data are consistent with a power-law dependence below $T \simeq 0.5$~K,
namely $\kappa/T = a + b T^c$, with a negligible residual linear term ($a \simeq 0$), a sub-linear power ($c \simeq 0.7$),
and a magnitude that varies by a factor 2 or so.
The strong similarity of $\kappa(T)$ reported here for dmit-131 and BEDT~\cite{Yamashita2009} is consistent with their very similar specific heat $C(T)$.
We attribute the magnitude and behavior of $\kappa(T)$ entirely to phonons as the sole carriers of heat,
which are scattered by low-energy spin excitations of the spin liquid state, which is typical of spin-liquid materials.

Our data on dmit-131 are in sharp disagreement with those of ref.~\cite{Yamashita2010}, where $\kappa(T)$ for dmit-131 is 50 times larger, with a quadratic $T$ dependence of $\kappa/T$ ($c = 2.0$),
extrapolating to a large residual linear term as $T \to 0$ ($a \equiv \kappa_0/T \simeq 2$~mW/K$^2$cm).
Our samples come from the same source as those used by Yamashita and colleagues, in one case (F1) grown using the very same starting material and in another case (G1) coming from the same growth batch as sample C of ref.~\cite{Yamashita2019}. Moreover, SEM measurements have shown that no micro-cracks are present in our samples, and XRD measurements have demonstrated that our samples - both prior and after our thermal conductivity measurements - are of similar quality as those employed by Yamashita. Consequently, it is unlikely that sample quality could explain the disagreement between our study and prior work~\cite{Yamashita2010,Yamashita2019}.
This leaves us today with no compelling, robust and reproducible experimental evidence of mobile gapless excitations in the low-temperature thermal conductivity of any candidate spin liquid.

We are aware that the group of Shiyan Li at Fudan University has obtained results similar to ours from their recent thermal conductivity study of dmit-131~\cite{Ni2019}.


{\it Note added.--}
Soon after our paper was first posted, Yamashita published a note~\cite{Yamashita2019} to propose a possible explanation for the huge discrepancy between our data and the data they reported in 2010~\cite{Yamashita2010}.
Importantly, whereas in 2010 only data from two samples (A and B) were reported, 
now in this note Kyoto data from an additional four dmit samples are presented (samples C, D, E and F).
We learn that $\kappa$ in their samples E and F is 50-100 times smaller than $\kappa$ in samples A and B.
This dramatic lack of reproducibility in measurements performed on nominally identical samples from a single source is cause for concern.


\begin{figure}[t!]
\includegraphics[scale=0.47]{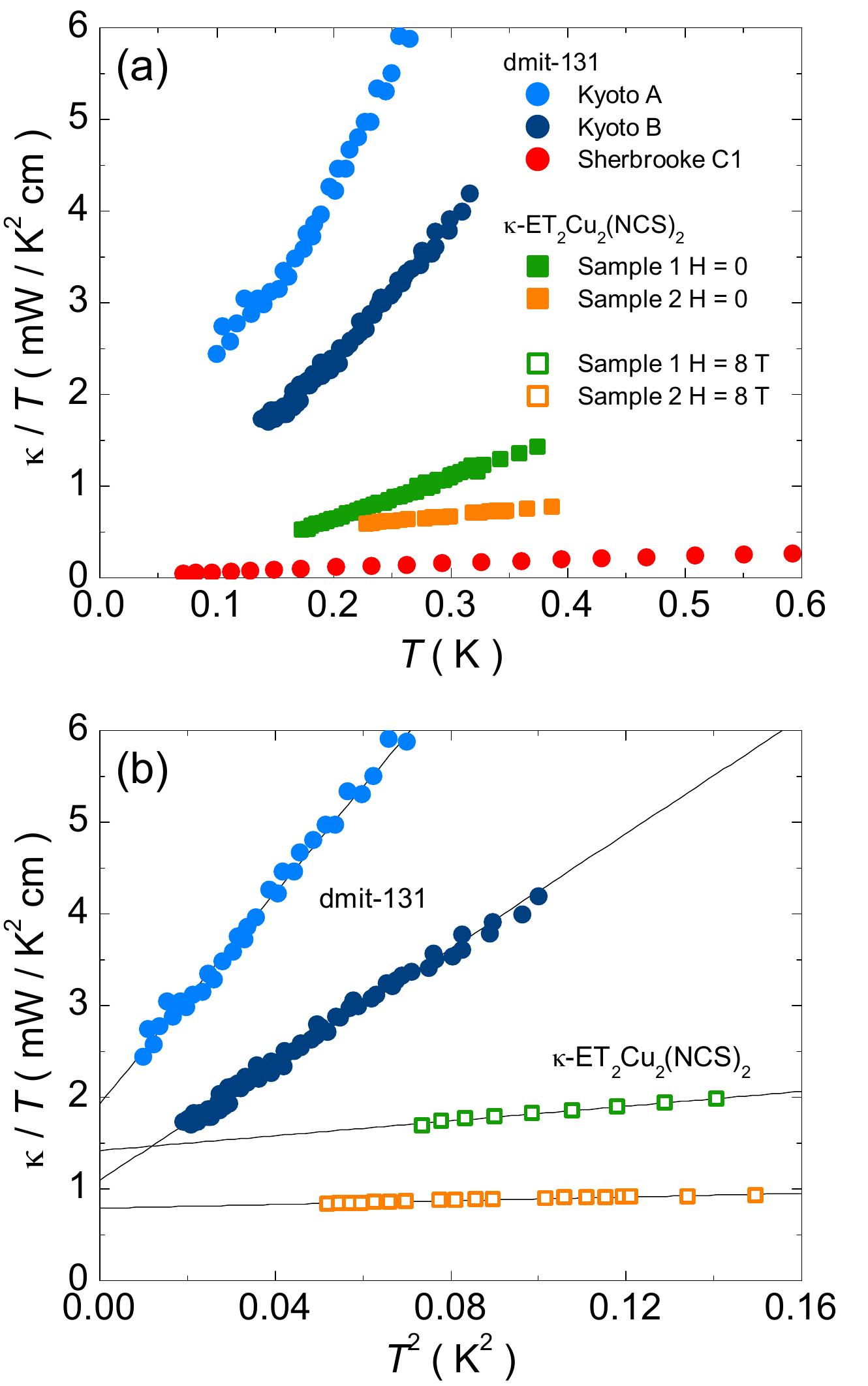}
\caption{
(a) Thermal conductivity $\kappa/T$ vs $T$ for dmit-131 (dots) as measured by the Kyoto group and at Sherbrooke, and for $\kappa$-ET$_2$Cu$_2$(NCS)$_2$ (squares) as reported in ref.~\cite{Belin1998}. In the top panel all data are in $H$ = 0, meaning that $\kappa$-ET$_2$Cu$_2$(NCS)$_2$ is in the superconducting state (full squares) where low-energy excitations are gapped and cannot scatter phonons. 
(b) Thermal conductivity $\kappa/T$ vs $T^2$ for Kyoto’s dmit-131 data and $\kappa$-ET$_2$Cu$_2$(NCS)$_2$ in its non-superconducting state in $H$ = 8 T (open squares), which reveals a comparable residual linear term, but a much smaller phonon slope in $\kappa$-ET$_2$Cu$_2$(NCS)$_2$. Lines are a fit of the data to $\kappa/T = a + b T^2$.
}
\label{FigB}
\end{figure}


Yamashita suggests that strong sample dependence could come from either structural domains that originate from the disordered cation, micro-cracks or different impurity levels. Impurities cannot be the 
cause of the huge difference between our sample F1 and the Kyoto samples A and B, since the same starting materials were used to grow these samples.
And our sample G1 and Yamashita's sample C (ref.~\cite{Yamashita2019}) are both coming from the same growth batch.
As mentioned in the Methods section, we performed careful XRD analysis, which revealed high crystal quality for all samples. This 
system has the degree of freedom of coordination of the ethyl group in the cation EtMe$_3$Sb$^+$. In the refinement of the structural model, 
however, the orientation is averaged due to space group constraints. Using high-intensity synchrotron radiation, orientational ordering at 
cryogenic temperatures can be clarified by analyzing the domain formation. However, the ethyl group ordering and domain formation 
associated therewith in this system could not be determined precisely due to strong effects of the thermal vibration even at 100K. It should 
be noted that there was no significant difference in the pattern of thermal diffuse scattering at 100 K between our crystals and the Kyoto 
samples.
Finally, scanning electron microscopy characterization of our 
samples does not reveal the presence of any micro-cracks.

Yamashita mentions that the high value of the phonon conductivity $\kappa_{\rm ph}$~measured in their `high-$\kappa$' samples
(A, B, C, D) is comparable to the value of $\kappa_{\rm ph}$ measured in other organic materials.
However, the comparison is not so simple because the two materials he mentions that were measured by other groups ($\kappa$-ET$_2$Cu(NCS)$_2$~\cite{Belin1998}
and $\lambda$-(BEDS)$_2$GaCl$_4$~\cite{Tanatar2002}) are not insulators but metals (and superconductors at low temperature).

In the superconducting state, electron states are gapped and so two things happen: the electronic contribution $\kappa_{\rm el}$ is reduced and $\kappa_{\rm ph}$ is increased, because phonons are less scattered by electrons. Applying a magnetic field to suppress superconductivity shows clearly that $\kappa_{\rm el}$ increases and $\kappa_{\rm ph}$ decreases. 
In Fig.~\ref{FigB}(a), we compare Kyoto’s dmit-131 data (samples A and B) with data on $\kappa$-ET$_2$Cu$_2$(NCS)$_2$ in the superconducting state, at $H$ = 0 (for the two samples reported in ref~\cite{Belin1998}). We see that the thermal conductivity of $\kappa$-ET$_2$Cu$_2$(NCS)$_2$ lies roughly halfway between the Kyoto data and our data on dmit-131. This shows that even when superconductivity gaps out the low-energy excitations in $\kappa$-ET$_2$Cu$_2$(NCS)$_2$, the phonons in that material do not conduct nearly as well as they appear to in Kyoto’s samples of dmit-131, a material whose low-energy excitations are not gapped (as seen in specific heat).

In Fig.~\ref{FigB}(b), we compare Kyoto’s dmit-131 data with data for $\kappa$-ET$_2$Cu$_2$(NCS)$_2$ now measured in the normal state, at $H$ = 8 T (ref~\cite{Belin1998}). We see that the residual linear term in the thermal conductivity of $\kappa$-ET$_2$Cu$_2$(NCS)$_2$ is comparable to that reported for Kyoto’s `high $\kappa$' dmit-131 samples (A, B, C, D). However, the phonon slope is now much smaller, by a factor ranging between 10 and 50. The low value of $\kappa_{\rm ph}$ in the normal state of $\kappa$-ET$_2$Cu$_2$(NCS)$_2$ is natural - this is what is observed in any metal. What appears unphysical is to have phonons propagating in a sea of zero-energy spin excitations and at the same time display a conductivity as if these excitations were absent, as Kyoto's dmit-131 data on samples A, B, C, and D suggest.

This raises a fundamental question about the Kyoto data on samples with a large residual linear term (A, B, C, D) -- 
comparable in fact to that measured in the metal $\kappa$-ET$_2$Cu(NCS)$_2$ 
(see lower panel of Fig.~\ref{FigB}).
How can it be that phonons in dmit are not scattered by the zero-energy spin excitations that give rise to the  residual linear term in the specific heat?
In a metal with electrons, $\kappa_{\rm ph}$ is strongly reduced because phonons are scattered by electrons -- as observed in $\kappa$-ET$_2$Cu(NCS)$_2$
(Fig.~\ref{FigB}).
In other spin-liquid materials, $\kappa_{\rm ph}$ is also strongly reduced (Fig.~\ref{FigA}).
We see no reason why phonons in dmit should be entirely decoupled from the low-energy spin excitations.


\section{Acknowledgements}

We thank S.~Fortier for his assistance with the experiments.
L.T. acknowledges support from the Canadian Institute for Advanced Research (CIFAR) as a CIFAR Fellow, and funding from 
the Institut Quantique, the Natural Sciences and Engineering Research Council of Canada (PIN:123817), 
the Fonds de Recherche du Qu\'ebec -- Nature et Technologies (FRQNT), 
the Canada Foundation for Innovation (CFI), 
and a Canada Research Chair.
This research was undertaken thanks in part to funding from the Canada First Research Excellence Fund.
This work was partially supported by the JSPS Grant-in-Aids for Scientific Research (S) (grant no. JP16H06346).


%

\end{document}